\documentclass[manuscript]{aastex}
\usepackage{graphicx}
\usepackage{epstopdf}
\usepackage{natbib}
\begin{document}

\title{Characterization of Active Main Belt Object P/2012 F5 (Gibbs): A Possible Impacted Asteroid}
\shorttitle{P/2012 F5}

\author{R. Stevenson*\altaffilmark{,1}, E. A. Kramer\altaffilmark{1,2}, J. M. Bauer\altaffilmark{1}, J. R. Masiero\altaffilmark{1}, A. K. Mainzer\altaffilmark{1}}

\altaffiltext{*}{To whom correspondence should be addressed: Rachel.A.Stevenson@jpl.nasa.gov}

\altaffiltext{1}{Jet Propulsion Laboratory, Caltech, 4800 Oak Grove Drive, Pasadena, CA 91109}
\altaffiltext{2}{Dept.\ of Physics, University of Central Florida, 4000 Central Florida Blvd.\,Orlando, FL 32816}

\begin{abstract}

In this work we characterize the recently discovered active main belt object P/2012 F5 (Gibbs), which was discovered with a dust trail $>$ 7$^{\prime}$ in length in the outer main belt, 7 months prior to aphelion.  We use optical imaging obtained on UT 2012 March 27 to analyze the central condensation and the long trail.  We find B-band and R-band apparent magnitudes of 20.96 $\pm$ 0.04 mag and 19.93 $\pm$ 0.02 mag, respectively, which give an upper limit on the radius of the nucleus of 2.1 km.  The geometric scattering cross-section of material in the trail was $\sim$ 4 $\times$ 10$^{8}$  m$^{2}$, corresponding to a mass of $\sim$ 5 $\times$ 10$^{7}$ kg.  Analysis of infrared images taken by the Wide-Field Infrared Survey Explorer in September 2010 reveals that the object was below the detection limit, suggesting that it was less active than it was during 2012, or possibly inactive, just 6 months after it passed through perihelion.  We set a 1-$\sigma$ upper limit on its radius during this time of 2.9 km.  P/2012 F5 (Gibbs) is dynamically stable in the outer main belt on timescales of $\sim$ 1 Gyr, pointing towards an asteroidal origin.  We find that the morphology of the ejected dust is consistent with it being produced by a single event that occurred on UT 2011 Jul.\ 7 $\pm$ 20 days, possibly as the result of a  collision with a small impactor.

\end{abstract}


\section{Introduction}

In recent years, a new class of objects with asteroid-like orbits but comet-like behavior has been identified within the main belt.  These objects occupy low-eccentricity, low-inclination orbits with Tisserand parameters ($T_{J}$) $>$ 3, placing them squarely in the asteroid regime.  However, they exhibit mass loss in a manner more akin to those observed around comets.  Though only a handful of these objects are known at this time, they cover a wide range of orbital space (2.29 AU $< a <$ 3.20 AU, 0.12 $< e <$ 0.34, 0$^{\circ}.2 < i <$ 21$^{\circ}$.4, where $a$, $e$, and $i$ represent semi-major axis, eccentricity, and inclination, respectively), their sizes vary by several orders of magnitude \citep{2012ApJ...747...49B}, and the morphologies observed include spherical comae, dust tails, and/or persistent debris trails.  The scientific community has yet to agree upon a name for such objects.  For this work, we find that the terms ``active asteroid'' \citep{2012AJ....143...66J}, ``activated asteroid'' \citep{2010IAUS..263..215L}, ``active main belt object'' \citep{2012ApJ...747...49B}, and the original definition of ``main belt comet" \citep{2006Sci...312..561H} all apply.  Potentially, depending on the driver of the observed activity, either the refined definition of main belt comet or ``disrupted asteroid'' \citep{2012ApJ...744....9H} may also apply.  For simplicity, in this work we choose to utilize the term ``active main belt object" (AMBO), which encompasses all objects with comet-like morphologies and main belt asteroidal orbits, regardless of the suspected driver of activity.

Studies have shown that most of the known AMBOs are dynamically stable on timescales longer than 100 Myr (e.g. \citealt{2012ApJ...744....9H,2012AJ....143..104H}), suggesting they are long-term residents of the main belt.  Conversely, 238P and P/2008 R1 are stable for only $\sim$ 20 - 30 Myr, respectively, and thus may be interlopers from elsewhere in the Solar System \citep{2009M&PS...44.1863H,2009AJ....137.4313J}.

One of the best characterized AMBOs, 133P/Elst-Pizzaro, has been active at multiple epochs, leading to the suggestion that the activity was driven by seasonal heating of an active area \citep{1996IAUC.6456....1E,2004AJ....127.2997H,2010MNRAS.403..363H}.
%One of the best characterized AMBOs, 133P/Elst-Pizarro, has been active at multiple epochs, most noticeably around perihelion \citep{1996IAUC.6456....1E,2004AJ....127.2997H,2010MNRAS.403..363H}, leading to the suggestion that the activity was driven by the sublimation of long-lived volatiles.  
A second AMBO, 238P/Read has also shown repeated activity, lending credence to the sublimation hypothesis \citep{2011ApJ...736L..18H}.  Their behavior mimics that of dynamical comets, which become active within a few AU of the Sun as volatile deposits are heated.  Work by \cite{2008ApJ...682..697S} suggests that sub-surface ice deposits could survive in the main belt for billions of years at depths of just a few meters.  Spectroscopic searches for gas emission lines have, to date, been unsuccessful but this may stem from the weakness of the outgassing (e.g. \citealt{2009AJ....137.4313J,2011A&A...532A..65L,2012ApJ...744....9H}).  Observations of 596 (Scheila) and P/2010 A2 (LINEAR) did not appear to fit with the model of mass loss driven by prolonged sublimation.  Instead, these asteroids displayed morphologies better explained by impulsive mass loss events, consistent with collisions \citep{2010Natur.467..817J,2010Natur.467..814S,2011ApJ...733L...4J}.

On UT 2012 Mar.\ 22.89, A.\ R.\ Gibbs reported the discovery of a new comet that appeared with a narrow dust trail greater than 7$^{\prime}$ in length.  Subsequent observations confirmed the discovery and the apparent morphology \citep{2012CBET.3069....1G}.  The object was given the cometary designation P/2012 F5 (Gibbs), hereafter referred to as P/2012 F5.  
%Preliminary orbital elements ($a$ = 3.05 AU, $e$ = 0.38, $i$ = 13$^{\circ}$.5) gave $T_{J}$ = 3.08, placing P/2012 F5 dynamically in the asteroid regime.  
On UT 2012 Jun.\ 21, ephemerides from the JPL HORIZONS System\footnote[2]{http://ssd.jpl.nasa.gov/?horizons} resulted in orbital elements of $a$ = 3.0038 AU, $e$ = 0.042, and $i$ = 9$^{\circ}$.739  (Table~\ref{table:orbels}), an orbital period of 5.21~yr, and a Tisserand parameter of $T_{J}$ = 3.23.  The corresponding 1-$\sigma$ uncertainties are 6.3 $\times$ 10$^{-5}$ AU, 8.4 $\times$ 10$^{-5}$, and 4$^{\circ}$.1 $\times$ 10$^{-4}$ for $a$, $e$, and $i$ respectively.  The comet-like appearance but asteroid-like orbit of P/2012 F5 make it the 9$^{th}$ known AMBO.

In this work, we characterize P/2012 F5 using optical imaging data obtained at Palomar Observatory just 3 days after its discovery.  We also use archived infrared images taken by the Wide-Field Infrared Survey Explorer (WISE) in September 2010 when the comet was 6 months post-perihelion.  Through aperture photometry, we constrain the size of the nucleus, and estimate the mass of dust within the trail.  We use the morphology and dynamical modeling of the trail to characterize the duration and onset time of the activity, as well as the properties of the dust grains emitted.  Finally, the dynamical stability of P/2012 F5 is investigated through the use of a symplectic integrator.  Our results are summarized in the conclusions.

\section{Observations and Data Reduction}

In this paper, we use optical groundbased observations and infrared spacebased observations to characterize P/2012 F5.  Table~\ref{table:obs} provides a summary of observations.

\subsection{Large Format Camera, Palomar Observatory}

We observed P/2012 F5 with the Large Format Camera (LFC) \citep{2000AAS...196.5209S} mounted on the 200\verb+"+ Hale telescope atop Mount Palomar on UT 2012 Mar.\ 27.  The array of six 2048 $\times$ 4096 pixel CCDs provided a 24$^{\prime}$ on a side field of view, with a resolution of 0$^{\prime\prime}$.35 pixel$^{-1}$ when using 2 $\times$ 2 binning.  A dithering pattern was invoked to provide coverage in the $\sim$ 15$^{\prime\prime}$ gaps between the chips.  We used B and R Bessel filters with central wavelengths and band widths of 4400 \AA\ and 1000 \AA, and 6300 \AA\ and 1200 \AA, respectively.  Two 180 s exposures and 5 90 s exposures were obtained with the B and R Bessel filters, respectively.  The object was tracked at non-sidereal rates as given by the JPL Horizons ephemeris.

The images were debiased and flattened using bias and twilight flatfield frames obtained on the night of the observations.  Amplifier glow impacted all of the images in the southern-most chip.  We investigated the extent of the contamination by comparing the median of a $\sim$ 114$^{\prime\prime}$ $\times$ 20$^{\prime\prime}$ box near the area of P/2012 F5's trail closest to the region of image most affected by the amplifier glow to the median of a background region far from both the amplifier and P/2012 F5.  We found that the amplifier glow had no effect on the R-band photometry, but that the amplifier glow contributed $\sim$ 0.03 mags to the background flux at the north-western end of the trail in the longer exposures obtained in B-band.

Photometric calibration was done using Landolt standards \citep{1992AJ....104..340L}.  Photometric uncertainties, including photon statistics, were found to be less than 0.05 mag.  The debiased, flattened images were median-combined to produce a deeper single image in each filter.  The stellar PSF full-width half-maximum (FWHM) was on the order of 1$^{\prime\prime}$.7 for the individual images.

\begin{figure}
\centering
\includegraphics[totalheight=8cm]{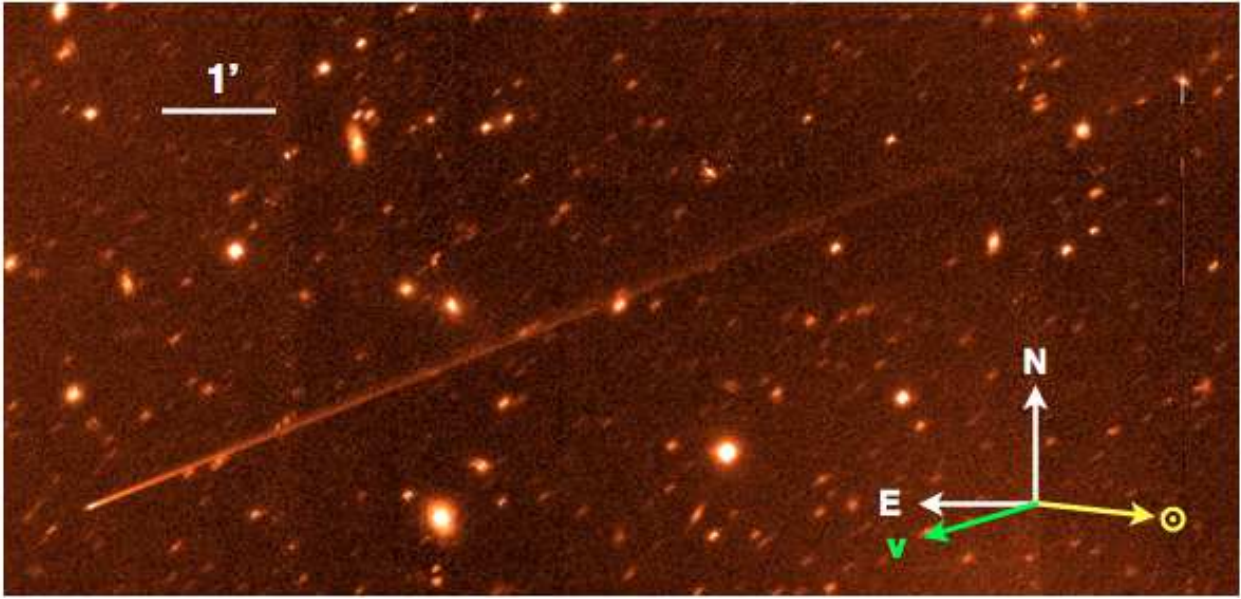}
\caption{P/2012 F5 as observed on UT 2012 Mar.\ 27 using the Large Format Camera mounted on the Hale 200" telescope atop Mount Palomar.  The image was created from a median stack of 5 90 s R-band exposures taken while tracking the motion of P/2012 F5.  The velocity vector, $v$, and the solar direction, $\odot$, are marked, as are the directions North and East.}
\label{fig:Rexp}
\end{figure}

The observations were obtained when P/2012 F5 was approximately 7 months pre-aphelion (true anomaly = 141$^{\circ}$) at a heliocentric distance of 3.10 AU.  The object appeared as a bright yet unresolved condensation in the south-east corner of the image with a long trail (Figure~\ref{fig:Rexp}).

\subsection{Wide-Field Infrared Survey Explorer}

We used archived data from the 40 cm WISE telescope to search pre-discovery images for signs of P/2012 F5.  The data were obtained between UT 2010 Sep.\ 22 and 24 after the cryogen was depleted in the secondary tank.  Consequently, only data in the 3 shortest wavelength bands (W1: 3.4 $\mu$m, W2: 4.6 $\mu$m, W3: 12 $\mu$m) were available.  The WISE field of view is 47$^{\prime} \times$ 47$^{\prime}$ with a pixel scale of 2$^{\prime\prime}$.75 pixel$^{-1}$.  Simultaneous exposures were taken in each band every 11 s, with nominal exposure times of 7.7 s in W1 and W2, and 1.1 s in W3 \citep{2010AJ....140.1868W,2011ApJ...731...53M}.   

Instrumental, photometric, and astrometric calibrations were performed by the ``first pass" scan/frame pipeline~\citep{2011wise.rept....1C}.  The apparent velocity of P/2012 F5 at the time of observation was 31$^{\prime\prime}$ hr$^{-1}$, corresponding to a drift across the frame of less than 0$^{\prime\prime}$.01 during the integrations.  Since this is far smaller than the pixel scale of each image, the effect of trailing was negligible.

We calculated the positional uncertainty of P/2012 F5 using information retrieved from the JPL Horizons Ephemeris service on UT 2012 Jun.\ 21.  The 3-$\sigma$ positional uncertainty was 2$^{\prime}$.5.  We thus rejected any frame in which the predicted position was less than 5$^{\prime}$ from the edge of the field of view.  This resulted in a total of 11 useful scans of the prediscovery field.  The individual images in each band were shifted to compensate for the motion of P/2012 F5 and were co-added using the ``A WISE Astronomical Image Co-adder" (AWAIC) algorithm \citep{2009ASPC..411...67M}, resulting in resampled images with pixel scales of 1$^{\prime\prime}$.0 pixel$^{-1}$.  The FWHM of the stacked images were  6$^{\prime\prime}$.1, 6$^{\prime\prime}$.4, and 6$^{\prime\prime}$.5 in bands W1, W2, and W3, respectively \citep{2010AJ....140.1868W,2011ApJ...731...53M}.  The 3-$\sigma$ uncertainty associated with the on-sky velocity of P/2012 F5 was small enough that the resulting deviation from its predicted position was several times smaller than the FWHM of each image.

The observation dates of the pre-discovery fields correspond to 6 months post-perihelion (true anomaly $\sim$ 38$^{\circ}$) when P/2012 F5 was at a heliocentric distance of 2.90 AU.

\section{Results}

\subsection{Nucleus Photometry at Optical Wavelengths}

Though the nucleus of P/2012 F5 is obscured by a dust coma, a bright condensation exists at the leading end of the trail.  We use aperture photometry to constrain its brightness with the goal of characterizing the source of the ejecta.  The aperture radius was selected to be $\sim$ 2.5 times larger than the FWHM of the images but small enough to focus on the material closest to the nucleus.  The aperture was centered on the opto-center of the condensation.  We note that the aperture contains non-negligible contamination from the coma and dust trail, and present the resulting cross-sections and radii as upper limits on the nucleus only.  Using a circular aperture with a radius of 2$^{\prime\prime}$.1 (3260 km as projected on the sky at the heliocentric distance of P/2012 F5 at the time of observation), we calculate apparent B-band and R-band magnitudes of m$_{B}$ = 20.96 $\pm$ 0.04 mag and m$_{R}$ = 19.93 $\pm$ 0.02 mag, respectively.  The B-R color of the near-nucleus region is 1.03 $\pm$ 0.04 mag, which is consistent with solar colors.

We correct for the observing geometry by converting the apparent magnitudes, $m_{B}$, $m_{R}$, to absolute magnitudes, $H_{B}$, $H_{R}$, assuming a phase coefficient, $\beta_{\alpha}$, of 0.04 mag deg$^{-1}$, as is common for both active and inactive comets \citep{2004come.book..223L}, as well as C-type asteroids, which dominate the outer main belt \citep{2000Icar..147...94B}.   We calculate a phase correction of 0.23 mag for the data obtained at Palomar.  The corresponding absolute magnitudes in B-band and R-band respectively are 16.66 $\pm$ 0.04 mag and 15.63 $\pm$ 0.02 mag.  The concept of absolute magnitude is not strictly valid for extended objects, given that an aperture of fixed angular size will include varying amounts of coma when the comet is observed at different heliocentric distances. 

\subsection{Photometry of the Trail}

We rotate the stacked images of P/2012 F5 so that the trail is aligned horizontally with the image axis with the head of the trail to the left.  We place a rectangular aperture around the trail, using the visible extent of the trail to set its boundaries.  The trail is not symmetrical along its breadth and extends farther north-east than it does south-west.  The photometric aperture, therefore, is set to extend 8$^{\prime\prime}$.75 (1.4 $\times$ 10$^{4}$ km as projected on the sky) in the north-east direction, and 5$^{\prime\prime}$.25 (8.1 $\times$ 10$^{3}$ km projected distance) to the south-west and centered on the opto-center of the trail along its length.  The aperture is then divided into 144 (B-band) or 148 (R-band) segments along the length of the trail, each having dimensions of 3$^{\prime\prime}$.5 $\times$ 14$^{\prime\prime}$.0.  The aperture used in B-band had a total size of 504$^{\prime\prime} \times$ 14$^{\prime\prime}$, while the R-band aperture had a size of 518$^{\prime\prime} \times$ 14$^{\prime\prime}$, corresponding to physical distances of 7.8 $\times$ 10$^{5}$ km by 2.2 $\times$ 10$^{4}$ km, and 8.0 $\times$ 10$^{5}$ km by 2.2 $\times$ 10$^{4}$ km at the heliocentric distance of P/2012 F5 at the time of observation.  Due to the large size of the trail and relatively small motion of P/2012 F5, numerous background objects contaminate the aperture.  We digitally remove the 4 worst offenders from the stacked B-band image, and the 6 worst from the stacked R-band image using the IRAF task IMEDIT, which replaces marked pixels with values interpolated from nearby sky regions.  We conservatively estimate that background objects remaining in the aperture contribute an uncertainty of $\sim$ 0.2 mag.  The conversion of photometric magnitudes to physical properties, such as mass of the dust, are strongly dependent on the assumed properties of the dust, many of which, such as particle size distribution, reflectivity, and density, are uncertain by more than 20\%.  We therefore consider the intrusion of background objects an additional source of error that is of order or less than other sources.  In order to estimate the sky background along the trail, we use a similar method to \cite{2004AJ....127.2997H,2010MNRAS.403..363H} and place rectangular sky background apertures of 3$^{\prime\prime}$.5 $\times$ 2$^{\prime\prime}$.8 (5.4 $\times$ 10$^{3}$ km by 4.3 $\times$ 10$^{3}$ km projected) directly above and below the apertures along the trail.  The sky background is computed as the median of the pixels within these sky apertures, and is subtracted from each pixel within the box apertures placed along the trail.

\begin{figure}
\centering
\includegraphics[totalheight=5cm]{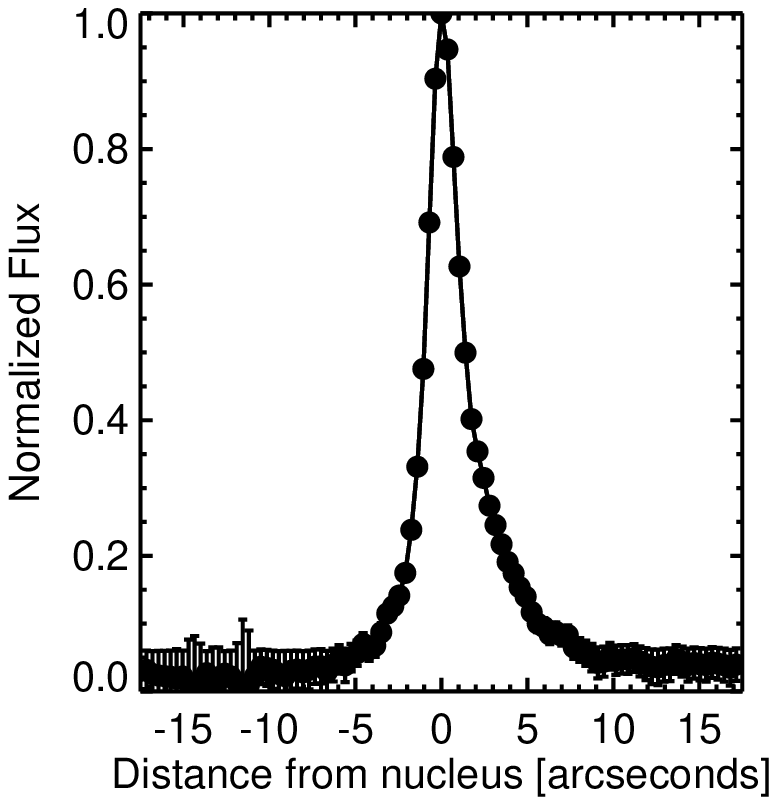}
\hskip 5pt
\includegraphics[totalheight=5cm]{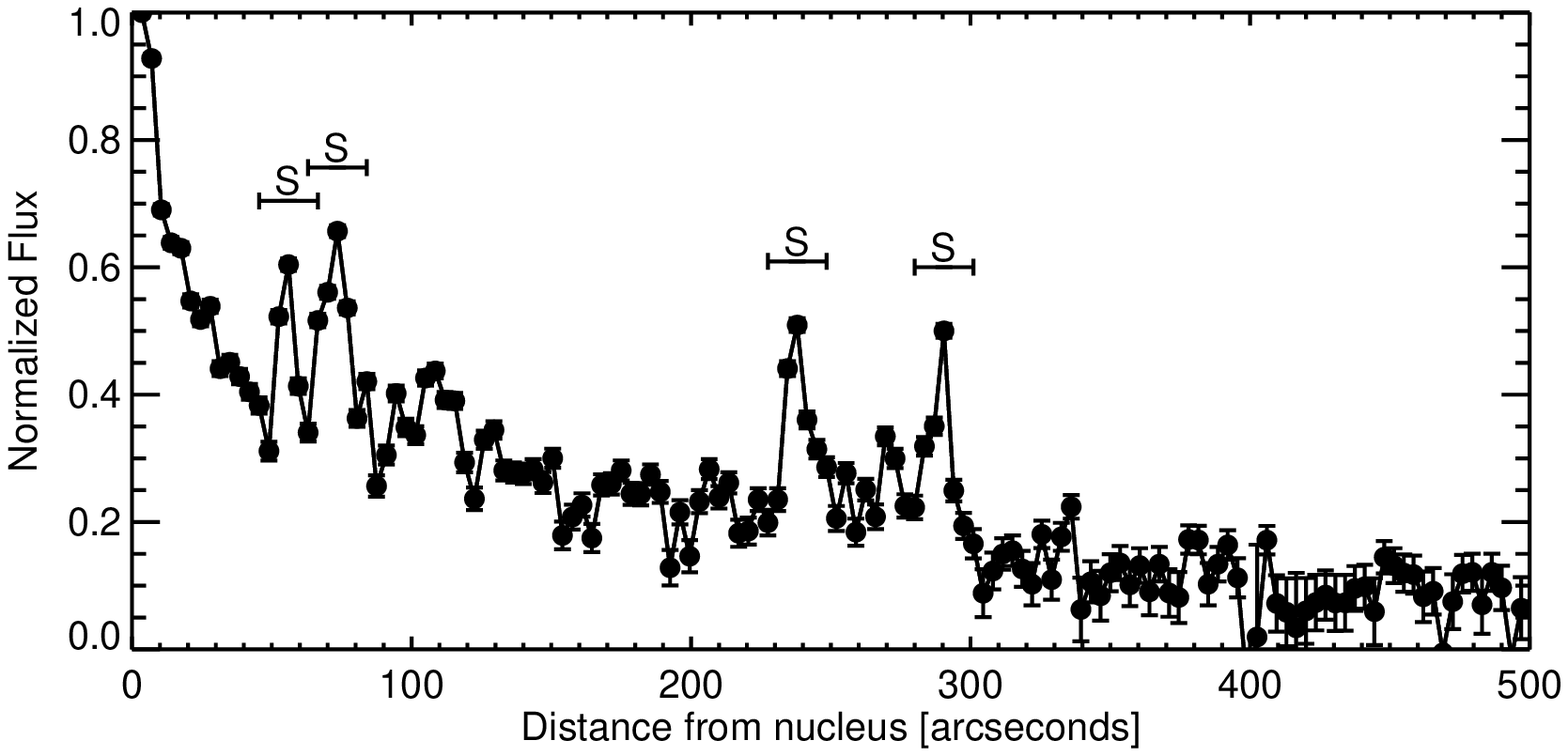}
\caption{R-band surface brightness profiles of P/2012 F5 along the breadth (left) and length (right) of the trail, calculated using a $\sim$ 500$^{\prime\prime} \times$ 14$^{\prime\prime}$ (8 $\times$ 10$^{5}$ km $\times$ 2 $\times$ 10$^{4}$ km projected) aperture.  Uncertainties due to photon statistics are plotted, though the scatter is increased through contributions from background objects.  Especially bright background stars are labeled with `S'.}
\label{fig:Rprof}
\end{figure}

Figure~\ref{fig:Rprof} shows the normalized brightness profiles of the trail, as measured along its breadth and length.  The cross-section through the trail's breadth is noticeably asymmetric, with the north-eastern edge (positive distance from the nucleus in Figure~\ref{fig:Rprof}) being more diffuse than the sharper south-western edge.  We note that many cometary dust tails and trails show asymmetries, perhaps most noticeably the trail of P/2010 A2 (LINEAR), which retained an unusual and highly asymmetric morphology for 9 months after undergoing an outburst in the main belt \citep{2010Natur.467..817J,2010ApJ...718L.132M,2010Natur.467..814S}.  Thus, this asymmetry may be a result of the intrinsic structure present in the ejected material shortly after the outburst.  The surface brightness decreases along the length of the trail with increasing distance from the nucleus, $d$, approximately as $d^{-0.4}$.  This is a somewhat shallower decrease than the relationship of $d^{-0.6}$ that was found for 133P/Elst-Pizarro as observed in 2002 by \cite{2004AJ....127.2997H}.
%This may be a result of a projection effect, fragmentation of grains within the tail, or non-isotropic ejection of material.   This may be a projection effect, or a result of fragmenting grains in the tail shedding smaller sized particles, which are then blown back in the anti-solar direction by radiation pressure.  The surface brightness along the length of the tail decreases rapidly with increasing distance from the nucleus.  
The material within the trail, as measured using the rectangular apertures previously described, has total magnitudes of 17.41 $\pm$ 0.20 mag and 16.25 $\pm$ 0.20 mag in B- and R-band, respectively.  We use 3$^{\prime\prime}$.5 $\times$ 14$^{\prime\prime}$.0 segments to measure the B - R color along the trail between the nucleo-centric distances of 14$^{\prime\prime}$ and 504$^{\prime\prime}$.  We find the median color of the segments to be 1.12 $\pm$ 0.28 mag,
%This gives a color of B - R = 1.16 $\pm$ 0.28 mag, 
consistent with both the color of the region at the head of the trail, and solar colors.  We note that the large uncertainty also renders it consistent with the color of dust around active comets (B - R $\sim$ 1.0 - 1.4 mag; \citealt{1986ApJ...310..937J,2004come.book..577K}).

\subsection{Constraints From WISE Infrared Photometry}

Using the stacked WISE images, an area of 600$^{\prime\prime} \times$ 600$^{\prime\prime}$ was searched for any PSF-like signal that exceeded a 1-$\sigma$ limit above the sky background using a 9$^{\prime\prime}$.0 radius aperture (1.8 $\times$ 10$^{4}$ km projected distance).  P/2012 F5 was not detected in any wavelength band by the WISE mission.  By characterizing the sky background in W3 (the most sensitive band to thermal dust emission) we set a 3-$\sigma$ detection limit on any potential objects within the field of 5 mJy, which corresponds to a lower limit on the absolute magnitude of material within the aperture of 15.2 mag.

\section{Analysis}

\subsection{Dust Morphology}
\label{sec:dustmorph}

The standard method for understanding the morphology of cometary dust tails and trails is the Finson-Probstein model \citep{1968ApJ...154..327F}. This model assumes that once cometary dust particles leave the surface, their motion is governed by two forces: solar gravity and solar radiation pressure. The particle motion can then be parameterized using the ratio, $\beta$ of these two forces:

\begin{equation}
\beta = \frac{F_{rad}}{F_{grav}}
\label{eq:beta}
\end{equation}

In physical units, this gives the ratio:

\begin{equation}
\beta = \frac{1.19 \times 10^{-3} Q_{pr}}{\rho_{d} d}
\label{eq:betaphys}
\end{equation}

where $Q_{pr}$ is the scattering efficiency for radiation pressure,  $\rho_{d}$ is the bulk density of the particle, $d$ is the particle diameter, and the factor of 1.19 $\times$ 10$^{-3}$ [kg m$^{-2}$] comes from multiplying all the constant values \citep{1968ApJ...154..327F}. 
For grain sizes of $d > \lambda$, $Q_{pr}$ $\sim$ 1 \citep{1979Icar...40....1B}; given the central wavelengths of the filters used, this condition holds for $d \gtrsim$ 0.5 $\mu$m.

$\beta$ is incorporated into the equation of motion in the following way:

\begin{equation}
\ddot{\vec{x}} + (1 - \beta) \frac{G M_{\odot}}{|\vec{x}|^{3}} \vec{x} = 0
\label{eq:motion}
\end{equation}

where $G$ is the universal gravitational constant, $M_{\odot}$ is the mass of the Sun, $\vec{x}$ is the vector position of the object. This is a simple equation of motion that can then be integrated for different values of $\beta$ to track the motion of particles with a particular $\beta$ value.

	The computations were carried out by creating a numerical integrator (based on the work of \citealt{1998ApJ...496..971L}) in the language Python which took in a set of $\beta$ values (0.0001 $< \beta <$ 3.0000), and integrated the motion of the dust particles over the designated time interval. This generated a set of points which can be shown as curves of particles with constant $\beta$ released at a range of times (syndynes) or curves of constant release date with a range of particle sizes (synchrones).

\begin{figure}
\centering \includegraphics[totalheight=12cm]{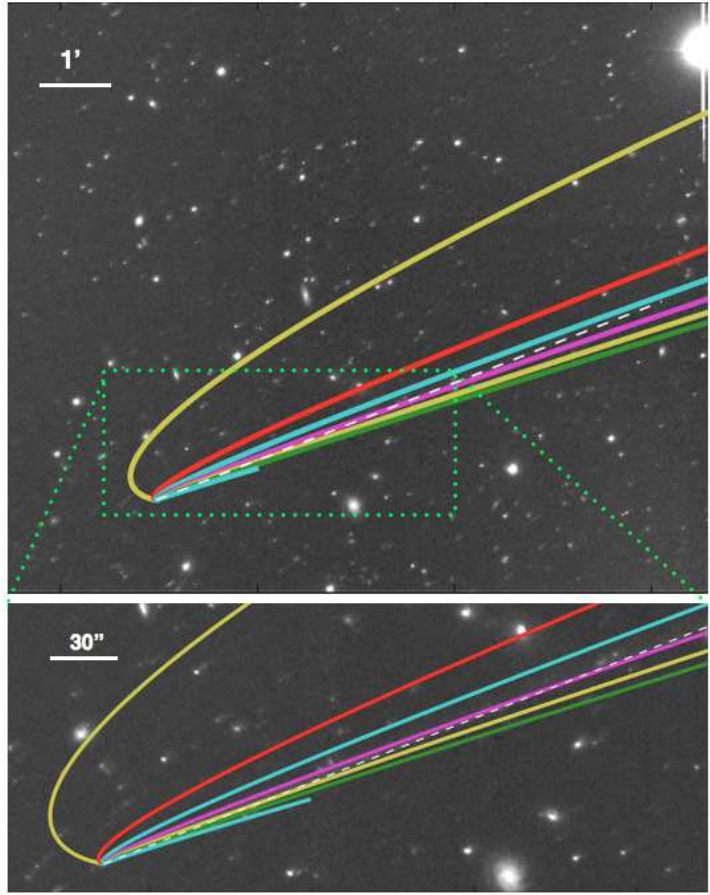}
\caption{Syndynes tracing particles with $\beta$ = 0.0003 - 3.0 (bottom to top, cyan through yellow) are plotted over a stacked R-band image of P/2012 F5 from UT 2012 Mar.\ 27.  The syndynes do not recreate the observed morphology and diverge from the trail (highlighted by the white dashed line) at large nucleo-centric distances.  The best-fitting syndyne ($\beta$ = 0.03, magenta) initially curves north of the trail then crosses to the south-west approximately 200$^{\prime\prime}$ (3.1 $\times$ 10$^{5}$ km projected) from the nucleus.  We find that synchrones provide a much better fit, suggesting the dust was released during a single event, rather than over a prolonged period.}
\label{fig:syndynes}
\end{figure}

	Figure~\ref{fig:syndynes} shows plots of the syndyne models plotted on top of the R-band data. At first look, the $\beta$ = 0.03 (shown in magenta) syndyne seems to model the trail (highlighted with a white dashed line) well. However, upon closer inspection it is clear that this syndyne initially is north-east of the trail, then curves south-west and crosses the trail approximately 200$^{\prime\prime}$ (projected distance: 3.1 $\times$ 10$^{5}$ km) from the nucleus. 

\begin{figure}
\centering \includegraphics[totalheight=7cm]{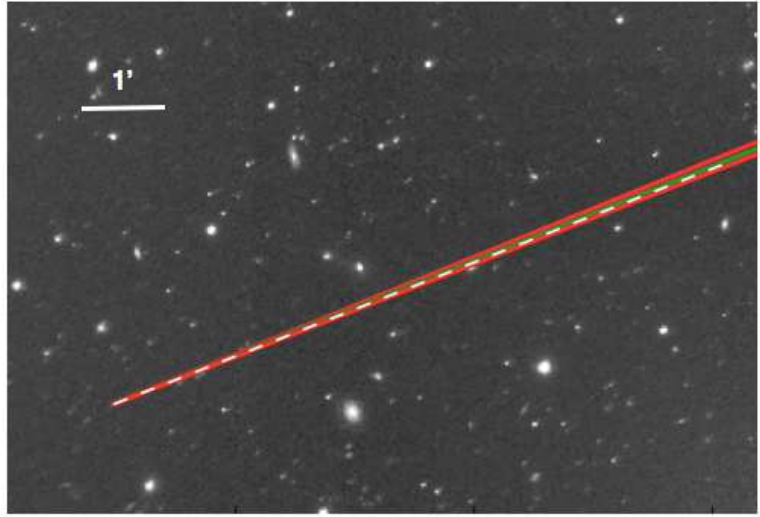}
\caption{The best-fit synchrone (green) is plotted on the stacked R-band image with additional synchrones representing the uncertainties (red).  The position and approximate length of the trail is marked by the dashed white line.  Particles represented by the best-fit synchrone must have been released 264 $\pm$ 20 days prior to the observation, which corresponds to an ejection date near UT 2011 Jul.\ 7.}
\label{fig:synchrones}
\end{figure}

	When the model is instead shown as synchrones (curves of constant particle release date, Figure~\ref{fig:synchrones}), the fit is greatly improved. 
	The particle emission date was constrained by modeling the width of the trail with a Gaussian, which yielded a half-width half-maximum of about 1$^{\prime\prime}$.05. We then calculated where the $\beta$ = 0.03 syndyne came within this distance from, and then crossed, the center of the trail. This revealed a best-fit synchrone of 264 $\pm$ 20 days, as shown as a green line in Figure~\ref{fig:synchrones}.  This corresponds to a particle ejection date of UT 2011 Jul.\ 7 $\pm$ 20 days when P/2012 F5 was at a heliocentric distance of 3.01 AU and a true anomaly of $\sim$ 94$^{\circ}$.
	
	We use the observed length of the trail to constrain the size of the particles, given that the distances travelled by dust grains since the ejection event are size-dependent.  We find that particles with radii of $\sim$ 20 $\mu$m could have traveled $\sim$ 8 $\times$ 10$^{5}$ km to the end of the observed trail in 264 days, setting a lower limit on the size of particles present.  Smaller particles could have traveled further but are not observed in our data, possibly because they were either not originally released or had a scattering cross-section that fell below our detection limit. 
	
\subsection{Physical Characterization of P/2012 F5 and its Activity}
\label{sec:tailphot}

The results from the aperture photometry can be used to constrain the physical properties of the nucleus and adjacent coma.  We use Equation~\ref{eq:sigma} to estimate the geometric cross-section of material, $\sigma$ [m$^{2}$] within the aperture.

\begin{equation}
p_{\lambda} \Phi_{\alpha} \sigma = 2.24 \times 10^{22} ~\pi ~r_{H}^{2} \Delta^{2} 10^{0.4 (m_{\odot\lambda} - m_{\lambda})}
\label{eq:sigma}
\end{equation}

where $p_{\lambda}$ is the geometric albedo, $m_{\odot\lambda}$ is the apparent magnitude of the Sun, and $m_{\lambda}$ is the apparent magnitude of the material within the aperture, each given for a specific broadband filter \citep{1991ASSL..167...19J}.  We assume the albedo of material is 0.04 in both B- and R-bands, which is broadly consistent with observations of AMBOs, outer main belt asteroids, and comets (e.g. \citealt{1986ApJ...310..937J,2004come.book..223L,2009ApJ...694L.111H,2011ApJ...741...68M,2012ApJ...747...49B}).  The apparent magnitudes of the Sun in B-band and R-band are -26.10 and -27.12, respectively \footnote[1]{\emph{http://mips.as.arizona.edu/$\sim$cnaw/sun.html}}.  The parameter $\Phi_{\alpha}$ is a function to correct for phase-angle-dependent variations in brightness and is calculated using Equation~\ref{eq:phasefunc}:

\begin{equation}
-2.5 log_{10} \Phi_{\alpha} = \alpha \beta_{\alpha}
\label{eq:phasefunc}
\end{equation}

where $\alpha$ [deg] is the phase angle and $\beta_{\alpha}$ [mag deg$^{-1}$] is the phase coefficient, already defined as having a value of 0.04 mag deg$^{-1}$.  

Considering the apparent magnitudes of material within a 2$^{\prime\prime}$.1 aperture, we estimate the geometric cross-section of the material to be $\sim$ 1.4 $\times$ 10$^{7}$ m$^{2}$ in B-band and $\sim$ 1.5 $\times$ 10$^{7}$ m$^{2}$ in R-band.  If the light was being reflected by an inactive, spherical body, the effective radius of the object would be $\sim$ 2.1 km.  This sets an upper limit on the size of P/2012 F5, though, given the extensive nature of the coma, the radius of the asteroid is probably on the order or less than 1 km.

We also use the non-detection of P/2012 F5 in the WISE data to set an upper limit on the size of a bare nucleus that could have been present during September 2010.  The upper limit on the flux from the W3 stacked image and the  NEATM model \citep{1998Icar..131..291H,2003Icar..166..116D,2011ApJ...736..100M} with a fixed beaming ($\eta$) parameter of 0.8 \citep{1986Icar...68..239L} and an assumed optical albedo of 0.04 $\pm$ 0.02 yielded a radius of $<$ 2.9 km, for a 1-$\sigma$ confidence interval.  

The upper limits on the size of P/2012 F5 are consistent with reported sizes of other AMBOs, which range in radius from 120 m to 113 km, with a median radius of 2.2 km.  It should be noted that all but one AMBOs have reported radii under 5 km.   

%The diameter upper limit is unremarkable for an asteroid in the outer main belt \citep{2011ApJ...741...68M}.  

To determine whether P/2012 F5 would have been observed by WISE if it displayed a similar brightness and morphology to its appearance in the Palomar images, we compare the lower limit on absolute magnitude, $H_{R,WISE}$ $\gtrsim$ 15.2, to that of P/2012 F5 during the Palomar observations.  We use a 11$^{\prime\prime}$.9 radius aperture (projected distance 1.8 $\times$ 10$^{4}$ km) to match the physical distance subtended by the 9$^{\prime\prime}$.0 radius aperture used with the WISE data.  Centering this aperture on the condensation observed in March 2012, we find an apparent R-band magnitude of 18.67 $\pm$ 0.02 mag, corresponding to an absolute magnitude of 14.37 $\pm$ 0.02 mag.  This is $\sim$ 0.8 mag brighter than the limit set by WISE, suggesting that P/2012 F5 would have been detectable by WISE in September 2010, had it displayed 2012-levels of activity.  We can thus say that the object must have had a different morphology when observed in 2010, and was possibly a bare asteroid.

We use the quantity $Af\rho$ as a measure of dust production as initially defined in \cite{1984AJ.....89..579A} and further discussed in \cite{1995Icar..118..223A}.  We adapt the initial formulation given in \cite{1984AJ.....89..579A} to give:
 
\begin{equation}
A f \rho = \rho \Bigg(\frac{2 \Delta r_{H}}{a \rho} \Bigg)^{2} 10^{0.4(m_{\odot \lambda} - m)}
\label{eq:afp}
\end{equation}

where $A$ is the geometric albedo, $f$ is a filling factor of the grains in the field of view, $\rho$ is the projected size of the aperture [cm], $\Delta$ is the geocentric distance [cm], $r_{H}$ is the heliocentric distance [AU]. $m$ is the phase-corrected magnitude of the material, while $m_{\odot \lambda}$ retains its previous definition.  The model assumes the coma follows simple radial outflow model, which is insufficient to explain the observed morphology of P/2012 F5.  We thus restrict ourselves to using photometric results from small apertures ($< 3^{\prime\prime}$) when calculating $Af\rho$.  Since the coma does not follow the assumed radial model where N($\rho$) $\propto$ $\rho$, where N($\rho$) is the number of dust grains within an aperture of size $\rho$, we find that the calculated values of $Af\rho$ are not independent of aperture size.  We find that the value of $Af\rho$ peaks at $\rho \sim$ the FWHM of the image, and subsequently decreases by $\sim$ 1 cm for every 1$^{\prime\prime}$ increase in aperture radius.  For an aperture of radius 2$^{\prime\prime}$.1 (a projected distance of 3260 km), $Af\rho$ = 11.0 cm and 11.1 cm in B-band and R-band, respectively.  The uncertainties are dominated by the morphological divergence of the data from the model of simple, symmetric outflow and are significant enough that the results from the different filters are consistent with each other.  This suggests that the grains are gray in color.

%\subsubsection{The Tail}
%\label{sec:tailphot}

Equation~\ref{eq:sigma} can also be used to estimate the geometric cross-section of material within the trail if the apparent magnitudes of the trail are substituted for $m_{\lambda}$.  We find $\sigma$ = 3.6 $\times$ 10$^{8}$ m$^{2}$ and 4.1 $\times$ 10$^{8}$ m$^{2}$ for the B- and R-band magnitudes, respectively.  As discussed in Section~\ref{sec:dustmorph}, the minimum particle size within the trail is $\sim$ 20 $\mu$m, while the largest dust grains close to the nucleus may be $>$ cm-sized.  By making the simplifying assumption that the bulk of the mass of the trail resides in spherical grains with radii of 100 $\mu$m and bulk densities of 1000 kg m$^{-3}$, we find that the mass of material within our apertures is $\sim$ 5 $\times$ 10$^{7}$ kg.  Taking the radius of P/2012 F5 to be $\sim$ 1 km with a bulk density of 1000 kg m$^{-3}$, the material in the trail amounts to $\sim$ 0.01\% of the asteroid by mass.  

\subsection{Dynamical Stability}

\begin{figure}
\centering \includegraphics[totalheight=4.2cm]{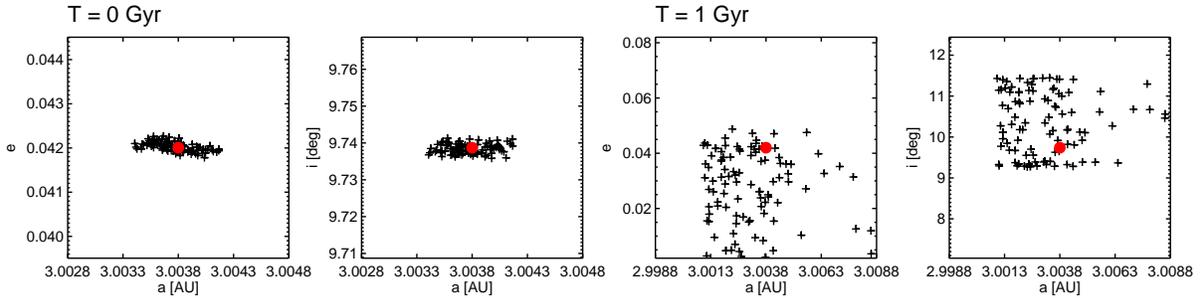}
\caption{Dynamical evolution of 100 clones (black crosses) that started with orbits within the quoted 1 $\sigma$ uncertainties of the orbit of P/2012 F5 (filled red circle).  After precessing the orbit for 1 Gyr, the particles have dispersed in orbital space but not significantly.  The object P/2012 F5 can thus be considered a dynamically stable main belt asteroid, and not an interloper from the outer Solar System.}
\label{fig:dynev}
\end{figure}

We next seek to constrain the origin of P/2012 F5 to test whether it it likely to have originated in the main belt, or to have been recently inserted from elsewhere in the Solar System.
Using numerical simulations to model test particle evolution, we have
investigated the stability of $100$ clones of P/2012 F5 on the
Gyr-timescale.  Clone starting positions were randomly assigned according to a Gaussian distribution with a FWHM equal to the 1-$\sigma$ uncertainties in the osculating elements of P/2012 F5 given by the JPL HORIZONS System.  
We integrated the orbits of each clone for $1~$Gyr
using the SWIFT\_RMSVY symplectic integrator \citep{1994Icar..108...18L,broz06} with a 25 day timestep,
and including the Venus, Earth, Mars, Jupiter, Saturn, Uranus, and Neptune as massive bodies
in addition to the Sun.  We ignore non-gravitational effects for the
purpose of this simulation as they are under-constrained.  No clones were lost from the system during the simulation and the orbital evolution of the clones is negligible over significant periods of time  (Figure~\ref{fig:dynev}), indicating that P/2012
F5 is dynamically stable and is not likely to have been recently implanted in the Main Belt.

%In order to test whether P/2012 F5 is a member of a dynamical family,
%we apply the Hierarchical Clustering Method (HCM) following
%\citet{1990AJ....100.2030Z,nesvornyPDS}.  HCM takes an initial object and a
%velocity window and accretes all asteroids that are within that given
%velocity cut onto the family.  This processing is repeated for all
%members of the cluster until no new objects can be added.  In order to
%test P/2012 F5, we need to assume that its currently observed
%osculating orbital elements are (1) accurate despite its small
%observational arc of $\sim40~$days and (2) equivalent to its proper
%orbital elements.  Under these assumptions, P/2012 F5 is only
%barely a marginal member of the Eos family - a 4400-member collisional family with an estimated age of 1.3$^{+0.15}_{-0.2}$ Gyr \citep{2006Icar..182...92V}.  It only becomes
%incorporated into the cluster at a velocity cut very close to the
%level where background objects begin to dominate the clustering, known
%as the quasi-random level.  Given this result and the assumptions
%required for this analysis we cannot yet conclusively determine family
%membership.  

\section{Discussion}

Dynamical modeling of the ejecta of P/2012 F5 suggests that the activity was driven by a one-time event, approximately 260 days before our observations.  This rules out continuous sublimation over a period of months, though a short, intense burst of sublimating material over a period of days could recreate the morphology observed.  Considering the possibility that the activity is driven by the sublimation of sub-surface volatiles, we examine the orbital location of P/2012 F5 during the WISE and Palomar observations.  P/2012 F5 was at a heliocentric distance of 2.90 AU and 6-months post-perihelion during the WISE observations, which revealed that the object was not displaying the activity observed at its discovery in early 2012.  Strangely, P/2012 F5 was discovered only months before reaching aphelion, at a time when its surface temperature is decreasing as its heliocentric distance increases.
However, the eccentricity of P/2012 F5 is only 0.04, corresponding to only a $\sim$ 6 K difference in surface temperature between perihelion and aphelion (assuming it absorbs and emits as an isothermal blackbody with an albedo of 0.04). We therefore do not expect P/2012 F5 to be active only near perihelion, if the activity is driven by a temperature-dependent process, such as sublimation.  It is difficult to reconcile the short period of activity with sublimation when dust tails, trails and/or comae observed around other AMBOs, such as 133P, 238P, and P/2010 R2, are consistent with prolonged generation over a period of several months or more \citep{2004AJ....127.2997H,2009AJ....137..157H,2012AJ....143..104H,2011ApJ...738L..16M}.  We note that if the estimated ejected mass lost through sublimation over a period of two weeks, the mass loss rate would be $\sim$ 25 kg s$^{-1}$, orders of magnitude higher than that observed for other AMBOs whose activity is suspected to be driven by sublimation \citep{2004AJ....127.2997H,2009AJ....137.4313J,2009AJ....137..157H}.  Observations of repeated activity on P/2012 F5 would strengthen the argument for sublimation-driven activity, especially if subsequent mass loss occurred at a similar orbital position to the 2011 outburst studied here. 

Smaller asteroids are subject to non-gravitational forces that can cause rapid rotational-spin up \citep{2000Icar..148....2R}.  If the rotational velocity of material on the surface exceeds the escape velocity, material may be ejected \citep{2012AJ....143...66J}.
The fate of ejecta is not well constrained, but work by \cite{2008Natur.454..188W} suggests that the material would be lifted into a low orbit where it may eventually accrete into a satellite.  This is supported by observations of asteroid 1999 KW4, which is rotating with a period close to the limit, has a binary companion and is presumed to be losing mass from its elongated equator \citep{2006Sci...314.1276O,2006Sci...314.1280S}.  Due to a lack of well-studied examples, it is unclear whether rotational breakup could recreate the morphology observed around P/2012 F5, or whether it would eject sufficient mass on a short timescale of just a few weeks.  Future measurements of the rotation period and size of P/2012 F5 could assist with gauging the likelihood of this scenario by calculating whether the nucleus is rotating rapidly enough to eject material.

The results of the dust modeling are also consistent with an impact that occurred $\sim$ 260 days prior to our observations.  If the activity observed around P/2012 F5 is due to a collision with a smaller asteroid, the mass of observed ejecta can be used to set an upper limit on the size of a possible impactor.  A mass of 5 $\times$ 10$^{7}$ kg, as calculated in Section~\ref{sec:tailphot}, would correspond to a sphere of radius 20 m, assuming a bulk density of 1000 kg m$^{-3}$.  As \cite{2010Natur.467..817J} note, the majority of the mass lost from a small body during an impact is from the primary body, rather than the impactor.  We thus conclude that the radius of an impactor would have been on the order of meters in size.  %We use equation~\ref{eq:tc} to estimate the collision timescale, $\tau_{c}$ [yr], for P/2012 F5:
The ejecta mass is comparable to estimates of the ejected mass around P/2010 A2 (3.7 $\times$ 10$^{7}$ kg, \citealt{2010Natur.467..814S}; 5 $\times$ 10$^{7}$ kg, \citealt{2010ApJ...718L.132M}; 6-60 $\times$ 10$^{7}$ kg, \citealt{2010Natur.467..817J}) and 596 (Scheila) (3 $\times$ 10$^{7}$ kg, \citealt{2012ApJ...744....9H}; 4 $\times$ 10$^{7}$ kg, \citealt{2011ApJ...733L...4J}; 1.5-4.9 $\times$ 10$^{8}$ kg, \citealt{2011ApJ...740L..11I}; 6 $\times$ 10$^{8}$ kg, \citealt{2011ApJ...733L...3B}; 2 $\times$ 10$^{10}$ kg, \citealt{2011ApJ...738..130M}), both of which are suspected to have undergone collisions in the main belt.  With so few examples, it is difficult to know if the similarity of the ejecta masses is coincidence or intrinsic to the nature of main belt asteroidal collisions.  

%\begin{equation}
%\tau_{c} = 4200~\Bigg(\frac{r_{p}}{1~\mathrm{m}}\Bigg)^{2.7} \Bigg(\frac{r}{1~\mathrm{km}}\Bigg)^{-2} 
%\label{eq:tc}
%\end{equation}

%where $r_{p}$ is the radius of the projectile [m] and $r$ is the radius of the primary [km] \citep{2012AJ....143...66J}.  Substituting $r_{p} =$ 1 m and $r =$ 1 km gives a collisional timescale of 4200 years for similar collisions within the main belt.  Given that there are $\sim$ 1.4 $\times$ 10$^{6}$ asteroids with radii $\geq$ 1 km, $\sim$ 330 similar events occur each year in the main belt.

%Future observations will likely provide greater insight into the cause of the observed activity.  The evolution of the tail may allow better constraints of the particle sizes and motions, while the nucleus may emerge from the coma, allowing measurement of its shape and size.

\section{Summary}

We have used optical and infrared data to characterize P/2012 F5, which appears to be a main belt asteroid undergoing mass loss as a result of an impact with a smaller body.  

\begin{enumerate}

\item A 1-$\sigma$ upper limit on the radius of the nucleus is set at 2.9 km using WISE observations from 2010.  A more sensitive upper limit of 2.1 km is set using R-band images from Palomar observatory.  The cross-section of material observed near the nucleus in 2012 is above the detection limit set by the WISE data, suggesting that the asteroid was either inactive or weakly active during September 2010.

\item Activity ejected $\sim$ 3 $\times$ 10$^{8}$ m$^{2}$ of material into the trail, spanning $>$ 7$^{\prime}$ (6.5 $\times$ 10$^{5}$ km projected distance) in space.  The minimum size of particles in the trail was $\sim$ 20 $\mu$m.

\item We find that syndynes do not fit the observed morphology, arguing against a period of continuous mass loss from the nucleus.  Instead, a single event on UT 2011 Jul.\ 7 $\pm$ 20 days can explain the observed shape of the trail.

\item Results from a symplectic integrator show the object to be dynamically stable and resident in the outer main belt on long timescales, suggesting P/2012 F5 is native to the main belt.

\item The observed behavior can be explained by a collision with a meter-sized impactor.  Further observations may reveal morphological evolution in the trail that may shed more light on the cause of the activity.

\end{enumerate}

\section{Acknowledgments}

This publication makes use of data products from the Wide-field Infrared Survey Explorer, which is a joint project of the University of California, Los Angeles, and the Jet Propulsion Laboratory/California Institute of Technology, funded by the National Aeronautics and Space Administration.  This work was based on observations obtained at the Hale Telescope, Palomar Observatory as part of a continuing collaboration between the California Institute of Technology, NASA/JPL, and Cornell University.  The authors thank Henry Hsieh for bringing the discovery of P/2012 F5 to their attention.  They also thank Kevin Rykoski, Carolyn Heffner, and Jean Mueller for assistance with the observations at Palomar Observatory.  RS and JM acknowledge support from the NASA Postdoctoral Fellowship Program.  EK was supported by the JPL Graduate Fellowship Program.  This research was funded in part by a grant from NASA through the Near Earth Object observing programs, for the NEOWISE project.  The supercomputer used in this investigation was provided by funding from the JPL Office of the Chief Information Officer.

\clearpage

\begin{deluxetable}{lccccccc}
\tabletypesize{\scriptsize}
\tablecaption{Orbital Elements of P/2012 F5}
\tablewidth{0pt}
\tablehead{
\colhead{Epoch} & \colhead{a [AU] \tablenotemark{1}} & \colhead{q [AU] \tablenotemark{2}} & \colhead{e \tablenotemark{3}} & \colhead{i [deg] \tablenotemark{4}} & \colhead{$\Omega$ [deg] \tablenotemark{5}} & \colhead{$\omega$ [deg] \tablenotemark{6}} & \colhead{M [deg] \tablenotemark{7}}}
\startdata
2012 Mar. 28.0 & 3.00 & 2.88 & 0.04 & 9.74 & 216.9 & 177.4 & 138.4\\
\enddata
\tablenotetext{1}{Semi-major Axis}
\tablenotetext{2}{Perihelion}
\tablenotetext{3}{Eccentricity}
\tablenotetext{4}{Inclination}
\tablenotetext{5}{Longitude of the ascending node}
\tablenotetext{6}{Argument of perihelion}
\tablenotetext{7}{Mean anomaly}
\label{table:orbels}
\end{deluxetable}

\begin{deluxetable}{lcccccccc}
\tabletypesize{\scriptsize}
\tablecaption{Journal of Observations}
\tablewidth{0pt}
\tablehead{
\colhead{Date (UT)} & \colhead{Observatory} & \colhead{Filter} & \colhead{N \tablenotemark{!}} & \colhead{Exposure time [s]} & \colhead{r$_{H}$ [AU] \tablenotemark{2}} & \colhead{$\Delta$ [AU] \tablenotemark{3}} & \colhead{$\alpha$ [deg] \tablenotemark{4}} & \colhead{Image scale [km pixel$^{-1}$]}}
\startdata
2010 Sep.\ 22-24 & WISE & W3 & 11 & 1.1 & 2.90 & 2.7 & 20.2 & 5400 \\
2012 Mar.\ 27.3 & Palomar & R & 5 & 90 & 3.10 & 2.1 & 5.8 & 535 \\
2012 Mar.\ 27.3 & Palomar & B & 3 & 180 & 3.10 & 2.1 & 5.8 & 535 \\
\enddata
\tablenotetext{1}{Number of exposures}
\tablenotetext{2}{Heliocentric distance}
\tablenotetext{3}{Geocentric distance}
\tablenotetext{4}{Phase angle}
\label{table:obs}
\end{deluxetable}

\end{document}